\documentclass[runningheads]{llncs}
\usepackage[T1]{fontenc}
\usepackage{graphicx}
\usepackage{xcolor}
\usepackage{array}
\usepackage{booktabs}
\usepackage{caption} 
\usepackage{multirow}
\usepackage{float}
\usepackage{arydshln}
\usepackage{longtable}
\usepackage{xltabular}
\usepackage{soul}

\usepackage[inline]{enumitem}

\begin{document}
\title{\includegraphics[width=0.08\textwidth]{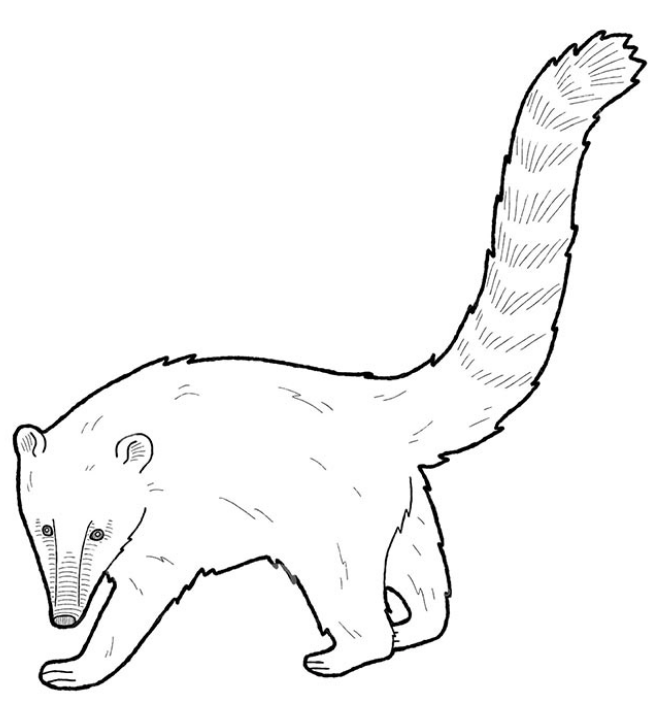}Quati: A Brazilian Portuguese Information Retrieval Dataset from Native Speakers}
\titlerunning{A Portuguese IR Dataset from Native Speakers}
%

\newcommand*\samethanks[1][\value{footnote}]{\footnotemark[#1]}

\author{Mirelle Bueno\thanks{Equal contribution.}\inst{1} \and
E. Seiti de Oliveira\samethanks\inst{1} \and
Rodrigo Nogueira\inst{1,2} \and
Roberto Lotufo\inst{1,3} \and
Jayr Pereira\inst{1,4}}

\authorrunning{Mirelle Bueno et al.}
%
\institute{University of Campinas \and
Maritaca AI \and
NeuralMind \and
Universidade Federal do Cariri}
%

\maketitle              

\setcounter{footnote}{0}

\begin{abstract}
    Despite Portuguese being one of the most spoken languages in the world, there is a lack of high-quality information retrieval datasets in that language. We present Quati,\footnote{We named our dataset after this South American mammal, whose foraging behavior represents the resolute search for resources.} a dataset specifically designed for the Brazilian Portuguese language. It comprises a collection of queries formulated by native speakers and a curated set of documents sourced from a selection of high-quality Brazilian Portuguese websites. These websites are frequented more likely by real users compared to those randomly scraped, ensuring a more representative and relevant corpus. To label the query--document pairs, we use a state-of-the-art LLM, which shows inter-annotator agreement levels comparable to human performance in our assessments. We provide a detailed description of our annotation methodology to enable others to create similar datasets for other languages, providing a cost-effective way of creating high-quality IR datasets with an arbitrary number of labeled documents per query.
    Finally, we evaluate a diverse range of open-source and commercial retrievers to serve as baseline systems. Quati is publicly available at \\
    \url{https://huggingface.co/datasets/unicamp-dl/quati} and all scripts at \\
    \url{https://github.com/unicamp-dl/quati}.
\end{abstract}

\section{Introduction}

The development of retrieval systems depends on high-quality evaluation datasets. Ideally, these datasets should contain queries and documents that native speakers wrote, so they capture the information needs and social-cultural aspects of the countries and communities that speak the language. This is to contrast with translated datasets, which, although in the same target language, represent the information needs and knowledge of a different society. Consequently, translated datasets may not effectively measure a retrieval system's ability in real-world scenarios involving native users.
In addition, test collections should have a high number of annotations whose documents were retrieved from a diverse pool of systems. This increases the chances that a dataset becomes ``reusable'', i.e., that they can reliably evaluate retrieval systems that did not contribute to the construction of the collection~\cite{voorhees2022can,voorhees2022too}.

Despite being one of the most widely spoken languages in the world, there is a scarcity of Information Retrieval (IR) datasets in Portuguese. Existing datasets such as REGIS~\cite{lima2021regis} and RCV2~\cite{lewis2004rcv1}\footnote{https://trec.nist.gov/data/reuters/reuters.html}, though valuable, fall short due to their limited size and specialized domains, such as geoscience and news. While translated datasets such as mMARCO~\cite{bonifacio2021mmarco} and mRobust04~\cite{jeronymo2022mrobust04} have helped to alleviate this issue, the use of automatic translations often represents the loss of socio-cultural characteristics of the target languages, thus, the evaluations may become biased from the source language~\cite{zhang2023m3exam}. Furthermore, the relevance annotation task is resource-intensive as it requires native speakers and a diverse set of retrievers to provide candidate query--passage pairs.


To address those issues we created Quati, a Brazilian Portuguese evaluation dataset, comprising human-written queries and a high-quality native corpus. Quati is created using a semi-automated pipeline, aiming to reduce the labeling cost barrier. We use a Large Language Model (LLM) to judge a passage's relevance for a given query, publishing a cost-effective pipeline to create an IR evaluation dataset with an arbitrary number of annotated passages per query.\footnote{The total cost for this dataset was U\$140.19 (0.03 per query-passage) for an average of 97.78 annotated passages per query.} To ensure the quality of the LLM annotations, we compare them with human annotations on a sample of query-passage pairs and confirmed a correlation of Cohen's Kappa 0.31. While this figure is below the 0.41 seen in human-human annotation agreement, it is consistent with the findings reported in the literature \cite{faggioli2023perspectives,thomas2023large,farzi2024exam} and it will likely increase as LLMs improve in quality. The usage of a modular semi-automated pipeline, allows the dataset construction method to be replicated to create high-quality IR datasets for other languages.


\section{Related Work}

Evaluation datasets are an important variable in the context of information retrieval as they expose the limitations of search systems and guide their development. However, most of the available datasets are in English, as is the case with MS MARCO~\cite{nguyen2016ms}. Works such MIRACL~\cite{zhang2023miracl}, mMARCO~\cite{bonifacio2021mmarco}, mRobust~\cite{jeronymo2022mrobust04}, Mr.Tydi ~\cite{clark2020tydi}, TREC CLIR ~\cite{schauble1998cross}, CLEF~\cite{peters2002importance}, NTCIR~\cite{sakai2021evaluating} and HC4 ~\cite{lawrie2022hc4} are efforts to develop datasets for other languages, but most are based on language translation to adapt English to the target languages, or do not include Portuguese.


The creation of datasets for IR is a resource-intensive task, particularly in the process of judging the relevance of documents. Recent endeavors have witnessed a shift towards leveraging LLMs to assess query--document relevance~\cite{zendel2024enhancing}. Faggioli et al.~\cite{faggioli2023perspectives} further underscored the potential of employing LLMs for automating the judgment of document relevance, thereby opening up promising avenues for exploration in this domain. Complementary evaluations conducted by Thomas et al. ~\cite{thomas2023large} demonstrated a significant correlation between human judgments and those made by the GPT-3.5-turbo model. 




Our work demonstrates the effectiveness of using the available LLMs (GPT-4) to perform the relevance judgment of Brazilian Portuguese (pt-BR) queries over a document collection originally created in the same target language, extracted from a large collection (ClueWeb22).

\section{Methodology}

\begin{figure}[ht]
\centering
\includegraphics[width=.9\textwidth]{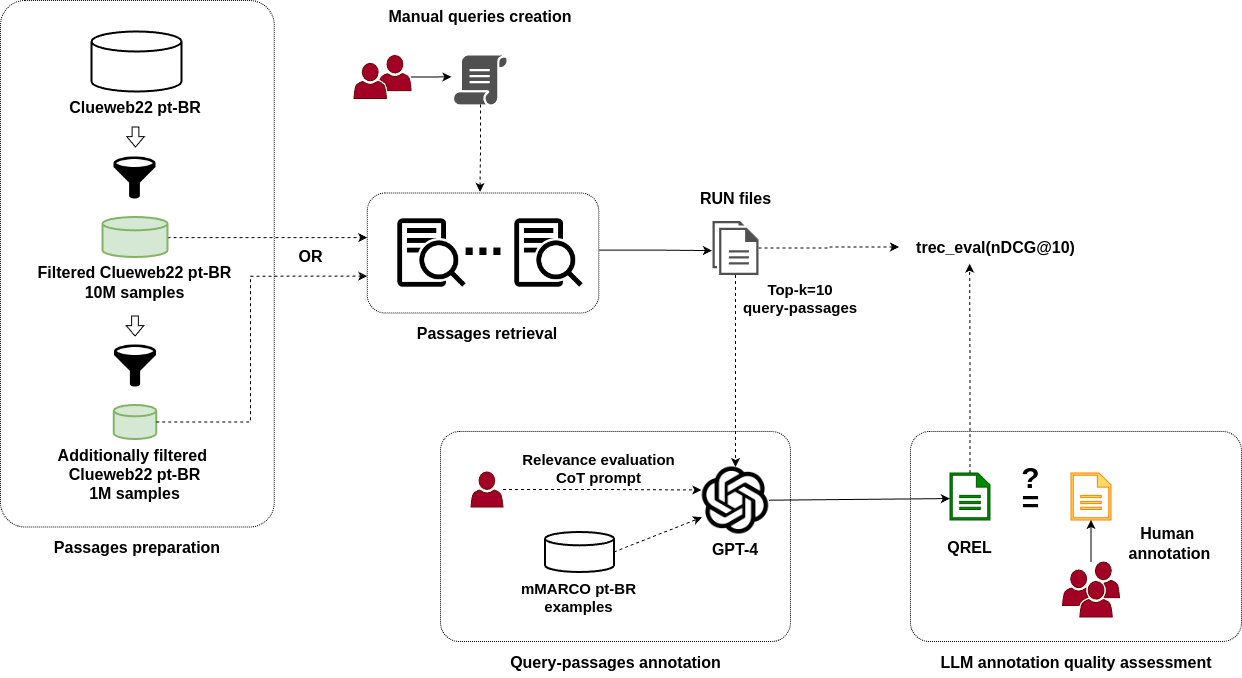}
\caption{\label{figure:pipeline}Proposed IR dataset creation methodology.}
\end{figure}

In this section, we describe the main steps to create the IR dataset in a semi-automatically manner, depicted in Figure \ref{figure:pipeline}. The required inputs are:

\begin{enumerate}
    \item A large corpus, originally written in the target language, from which we extract the passages to compose our IR dataset.

    \item A set of test queries, manually created to represent the information needs of native speakers.
\end{enumerate}

With those inputs we propose a pipeline to create the final IR dataset, aiming for a cost-effective creation of a high-quality evaluation set with an arbitrary number of passages evaluated per query.

\subsection{Passages preparation}

The passages preparation is composed by the following steps:
\begin{enumerate}
  \item Data collection;
  \item URL filtering;
  \item Document segmentation into passages; and
  \item Separation into large and small versions.
\end{enumerate}

\noindent\textbf{Data collection:} We used the Portuguese subset of ClueWeb22~\cite{overwijk2022clueweb22} category B, which includes 4.1 million web pages more likely to be visited according to Bing algorithms during the first half of 2022 \cite{overwijk2022clueweb22}. Our goal was to consider up-to-date content, not focusing on any particular domain.

\noindent\textbf{URL filtering:} Since our goal is to create a \textit{Brazilian} Portuguese dataset, we excluded any documents from our dataset whose URLs' domain ended with ``.pt'' as the language style in those documents might differ significantly from that used in Brazilian Portuguese web pages. Alternatively, we could have chosen to retain only ``.br'' URLs, but this would have excluded Brazilian Portuguese URLs that do not use this suffix, such as those used in Wikipedia.
Although the combined population of other Portuguese-speaking countries, such as Angola and Mozambique, has three times more native speakers than Portugal, their presence in ClueWeb, as measured by the number of .ao and .mz URLs, was much smaller than Portugal's. Therefore, we did not remove these URLs. Additionally, as the language identification algorithm used to construct ClueWeb22 may not be completely accurate, we used FastText \cite{joulin2016bag,joulin2016fasttext} as a secondary language verification method to ensure that only Portuguese documents were included in our corpus.

\noindent\textbf{Document segmentation into passages:} Following language verification, we segmented the documents into approximately 1,000-character segments and assessed the percentage of line breaks (\texttt{\textbackslash n}) occurrences within each segment, removing segments with more than 20\%. This criterion was used to increase the probability of retaining segments predominantly composed of natural language text.

\noindent\textbf{Separation into large and small versions:} Finally, we collected a total of 20 million segments. From this set, we randomly selected 10 million segments to be the passages in our corpus, creating a large, but still manageable, dataset of more than 11 GB of size. A second dataset was built from the first one applying additional filtering rules taken from the MassiveWeb Corpus \cite{rae2021scaling} and sampling only 1 million segments --- the goal was to create a smaller and higher-quality dataset that would facilitate experimentation with embedding models, as encoding the original 10 million segments can be computationally expensive.

\subsection{Manual queries creation}

We employed human-created queries for the evaluation dataset, looking for high-quality questions that resemble common information needs from a diverse corpus, created by native speakers of the target language. We created a total of 200 test queries considering two different approaches:

\begin{enumerate}

\item \textbf{Taxonomy-guided corpus-agnostic questions:}

We proposed the following taxonomy of themes, and questions characteristics to guide the first 100 queries creation in a corpus-agnostic fashion, i.e., without knowing in advance if the corpus would necessarily have good answers for them:

\begin{itemize}
    \item Themes: \textbf{Geography, Politics, Economy, Culture, Culinary, Tourism, Leisure, Sports}.
    \item Scope: \textbf{General}~(exploring a broad theme or subject) or \textbf{Specific}~(exploring a narrow theme or subject).
    \item Type: \textbf{Opinion}~(asking for an opinion about something) or \textbf{Factual}~(asking for a fact or data which has little dependency on one's opinion).
\end{itemize}

Two Brazilian Portuguese native speakers from the research team created those initial 100 queries.

\item \textbf{Questions from documents:}

Another 100 queries were created making sure the corpus included at least one good answer to the query: we sampled 100 passages we prepared from the original corpus and created one query regarding those subjects. That was performed by another member of our research team, also a Brazilian Portuguese native speaker.

\end{enumerate}

\subsection{Passages retrieval}\label{passages_retrieval}

Once we have the passages and the queries, the next essential step of our method is the passage retrieval. The aim of this step is to build a list of query--passages to annotate.
Ideally, we would like to have the relevance scores for each query for the entire corpus. As that is prohibitive in terms of annotation cost, an alternative is to select the top-k passages returned by multiple IR systems. It is assumed that the diversity of their results will enable the collection of a variety of passages which, together, will constitute a robust evaluation dataset.

We selected a mix of strong IR systems, aiming to get the most relevant passages within their top-k lists, as well as weaker systems, in order to also include less relevant passages:
\begin{itemize}
    \item \textbf{BM25}: as strong baseline for retrieval; we use the Pyserini implementation of BM25~\cite{Lin_etal_SIGIR2021_Pyserini}. We change the language to Portuguese at both index creation and retrieval.
    
    \item \textbf{BM25 + mT5-XL}: a two-stage pipeline with BM25 as the first stage, followed by mT5-XL (3.7 billion parameters) \cite{xue2020mt5} as a reranker.

    \item \textbf{BM25 + E5-large}: another two-stage pipeline with BM25 as first stage, and E5-large \cite{wang2022text}\footnote{https://huggingface.co/intfloat/multilingual-e5-large} as a reranker.

    \item \textbf{E5-large} and \textbf{E5-base}: directly using E5 variants as dense retrievers.

    \item \textbf{ColBERT-X}~\cite{nair2022transfer}, a multilingual ColBERT-v1 fine-tuned in Brazilian Portuguese subset of mMARCO. We indexed the 10M and 1M corpora using 48-dimension embeddings, to reduce CPU RAM requirements at retrieval time.

    \item \textbf{SPLADE v2}: a learned sparse retriever \cite{formal2021splade} fine-tuned on Brazilian Portuguese subset of mMARCO.
    
    \item \textbf{SPLADE v2 + mT5-XL}: a two-stage pipeline using SPLADE v2 followed by mT5 reranking.
\end{itemize}

For all two-stage pipelines, the second stage reranks the top 1,000 passages retrieved by the first stage. In addition to the retrieval system above, we also use Reciprocal Ranking Fusion (RRF) \cite{cormack2009reciprocal} to increase the retrieved documents diversity, using the following combinations:

\begin{itemize}
    \item E5-large + ColBERT-X
    \item E5-large + SPLADE v2
    \item E5-large + BM25 + mT5-XL
\end{itemize}

We also include the following commercial embedding models:

\begin{itemize}
    \item \textbf{text-embedding-ada-002}: a 1536-dimension OpenAI embedding model released on 2022\footnote{https://openai.com/blog/new-and-improved-embedding-model}, using FAISS~\cite{johnson2019billion} for dense vectors search.

    \item \textbf{text-embedding-3-small}: an improved 1536-dimension OpenAI embedding model released on 2024\footnote{https://openai.com/blog/new-embedding-models-and-api-updates}, also using FAISS for dense vectors search. As these embedding dimensions are hierarchical, we also performed the retrieval using only the first 768 dimensions --- that was identified as text-embedding-3-small half.

    \item \textbf{text-embedding-3-large}: a 3072-dimension OpenAI embedding model, also released in 2024, once again using FAISS for dense vectors search.
\end{itemize} 

To evaluate the diversity of retrieved passages, we counted the query--passage combinations exclusively returned by each IR system, which should be a number from 0 to 500, 0 meaning the query--passages returned by a particular IR system were also returned by another IR system, and higher counting indicates the individual system contribution to the dataset diversity is also high. Although we look for diversity, there should be a balance: 10 different IR systems, 50 queries, and 10 passages per query, we could have reached a total of 5,000 different query--passage combinations if all systems returned exclusive passages, which would indicate the systems could not agree on which are the most relevant passages per query. Consequently, it is likely that passages retrieved by future IR systems would not be represented in our labeled dataset, and the annotations would not be effective for their evaluation. In other words, this outcome would them limit the generalizability of our dataset to new retrievers.

\subsection{Query--passages annotation}

The final step of the query annotation is to use an LLM to label the retrieved passages' relevance for each query. 
We selected the top-k=10 passages for a sample of 50 queries using all the retrieval systems considered on both the 10M and 1M corpora and sent them for LLM evaluation. For that, we created an LLM prompt using a few-shot Chain-of-Thought (CoT) approach \cite{wei2022CoT}, adopting the TREC 2021 Deep Learning track 4-score relevance annotation \cite{craswell2021trec}:

\begin{enumerate}
    \item \textbf{Irrelevant}: denoting that the passage is outside the scope of the question.
    \item \textbf{Relevant}: indicating that the passage pertains to the question's topic but does not provide a direct answer.
    \item \textbf{Highly relevant}: the passage answers the question, but lacks in clarity or includes unrelated information.
    \item \textbf{Perfectly relevant}: signifying that the passage answers the question with clarity and precision.
\end{enumerate}

We selected OpenAI GPT-4 --- gpt-4-1106-preview	model\footnote{https://openai.com/blog/new-models-and-developer-products-announced-at-devday} --- as the annotator. Due to its cost, we sampled 50 queries from the full set of 200 examples. Then, we asked the LLM to label only the top-10 retrieved passages of each IR system for each query.

We used a CoT prompt with two in-context examples selected from the mMARCO pt-BR dataset \cite{bonifacio2021mmarco}. The prompt was written in Brazilian Portuguese, including the task explanation, and the CoT examples to instruct LLM to produce the 4-score passage relevance value for a given query. The final evaluation was requested in JSON format to simplify the LLM response parsing process --- our preliminary tests indicated the LLM responded better to a formal specification. We interactively built the prompt, manually verifying the evaluation results on a limited set of questions sampled from the same mMARCO pt-BR dataset. The final prompt version can be found online.\footnote{https://github.com/unicamp-dl/quati/blob/main/prompt.md}

\section{Experiments}

In this section, we describe how we evaluated the overall quality of the LLM query--passages annotations, the central part of our semi-automated pipeline for IR datasets creation.

\subsection{LLM annotation quality assessment}

We assess the quality of our LLM as an annotator by comparing its query--passage relevance scores with those provided by human annotators. Due to the time and effort required, this process was conducted on a subset of 24 queries, rather than the full set of 50 queries. Three non-expert annotators evaluated the top-10 passages returned by the BM25 + mT5 IR system using the same TREC-DL 2021 4-score grading system, already described in the previous section. The three researchers used the Doccano~\cite{doccano} system for annotating the 240 query--passage combinations. The agreement among the query-passage relevance annotations generated by the LLM and humans was measured using Cohen Kappa, Pearson, and Spearman correlation coefficients.

Following Thomas et al.~\cite{thomas2023large}, we consider the annotators to be in between the ``professional assessors'' and the ``crowd workers'' level. Although they cannot be considered ``subject-matter experts'' --- specially because the questions were broad in an open document basis --- the selected human annotators knew how the annotations are used and participated in trial sessions to clarify each score's meaning, creating an experience that would surpass the regular crowd worker one.

\subsection{Retrieval systems evaluation}

We used the LLM annotated query--passages to evaluate the IR systems effectiveness in the 10M and 1M Quati datasets. We simply compute the nDCG@10 metric over the IR systems runs we used to build the passages set we then sent for LLM annotations (see section \ref{passages_retrieval}. Besides establishing a baseline for a variety of IR systems, this experiments also indirectly assess the overall quality of Quati validation dataset: by verifying different effectiveness for already published IR systems, we validate Quati potential to indeed assess such systems --- the opposite scenario would be the IR systems showing very similar results over our validation dataset.

\section{Results and Discussion}

\subsection{Annotated passages variability}

Table \ref{tab:retrievals_exclusive_query_passages} depicts that count per IR system, indicating a range from 29 to 362 query--passage combinations exclusively returned by a single system. On average, each system returned 28.85\% of new passages, and from the total 4,889 evaluated query--passages, 61.96\% (3029) were returned by a single system. This suggests that our pool of retrieval systems is diverse, contrasting with a hypothetical scenario where all systems retrieve the same passages. As shown in Table \ref{tab:query_passages_score_stats}, the IR systems were able to retrieve a diversity set of query--passages, including ``perfectly relevant'' (score=3) ones; also, that diversity increased for less relevant passages, indicating the systems agreed more as the passage relevance increased, showing agreement and diversity at the same time.

\begin{table*}
    \centering
    \renewcommand{\arraystretch}{1.2}
    \begin{tabular}{>{\raggedright}p{7.0cm}>{\centering\arraybackslash}p{3.0cm}>{\centering\arraybackslash}p{2.0cm}}
        \textbf{Retrieval System} & \textbf{Single--system query--passages} & \textbf{\%}\\
        \midrule
        \textbf{1M passages} & & \\
        BM25 & 253 & 50.6\\
        ColBERT-X mMARCO pt-BR & 195 & 39.0\\
        text-embedding-ada-002 & 137 & 27.4\\
        text-embedding-3-large& 121 & 24.2\\
        BM25 + E5-large & 120 & 24.0\\
        BM25 + mT5-XL & 93 & 18.6\\
        text-embedding-3-small half & 45 & 9.0\\
        text-embedding-3-small & 31 & 6.2\\
        \midrule
        \textbf{10M passages} & & \\
        E5-base & 262 & 52.4\\
        BM25 & 248 & 49.6\\
        SPLADE v2 pt-BR & 151 & 30.2\\
        E5-large & 122 & 24.4\\
        ColBERT-X mMARCO pt-BR & 115 & 23.0\\
        BM25 + E5-large & 115 & 23.0\\
        SPLADE v2 pt-BR + mT5-XL & 86 & 17.2\\
        BM25 + mT5-XL & 60 & 12.0\\
        E5-large + ColBERT-X mMARCO pt-BR RRF & 32 & 6.4\\
        E5-large + SPLADE v2 pt-BR RRF & 29 & 5.8\\
        \midrule
        Others & 814 & 54.27\\
        \midrule
        \textbf{Single system query--passages total} & 3029 & 61.96\\
        \textbf{Union of all systems query--passages} & 4889\\
        \bottomrule
    \end{tabular}
    \caption{The single--system query--passages column correspond to the ones exclusively returned only by the corresponding retrieval system. The percentage for each system refers to the total of 500 query-passages returned by each system; for the single system total, the percentage refers to the total (union) of evaluated passages. ``Others'' includes the results of 3 runs which had data preparation issues; since the LLM annotated passages were valid, we kept them in the final dataset.} \label{tab:retrievals_exclusive_query_passages}
\end{table*}

\begin{table*}
    \centering
    \renewcommand{\arraystretch}{1.2}
    \begin{tabular}{>{\centering}p{1.8cm}>{\centering\arraybackslash}p{2.6cm}>{\centering\arraybackslash}p{3.0cm}>{\centering\arraybackslash}p{1.5cm}}
        \textbf{Score} & \textbf{All query--passages} & \textbf{Single-system query--passages} & \textbf{\%}\\
        \midrule
        0 & 2489 & 1839 & 73.89\\
        1 & 985 & 586 & 59.49\\
        2 & 759 & 375 & 49.41\\
        3 & 656 & 229 & 34.91\\
        \midrule
        \raggedleft \textbf{Relevant} & 2400 & 1190 & 49.58\\
        \raggedleft \textbf{Total} & 4889 & 3029 & 61.96\\
        \bottomrule
    \end{tabular}
    \caption{Query--passage relevance score counts. Comparing the ones returned by a single-system, it is possible to verify the systems agreed more, returning the same passages per query, as the relevance score increases, indicating a balance between the query--passages diversity and agreement. The relevant count includes passages from scores 1 to 3.}
    \label{tab:query_passages_score_stats}
\end{table*}

\subsection{LLM annotations quality is aligned with crowd workers}

Table \ref{tab:cohen_4_score} shows the Cohen's Kappa correlations for the human and LLM annotations, computed for all the 240 query-passage combinations at once. The average Cohen's Kappa of 0.31 is aligned with what has already been reported in the literature: Faggioli et al. reported Cohen's Kappa of 0.26 for GPT-3.5~\cite{faggioli2023perspectives}, and Thomas el al. reported Cohen's Kappa ranging from 0.20 to 0.64~\cite{thomas2023large}, depending on the evaluation prompt used on GPT-4. Our human annotators' mean Cohen's Kappa correlation is 0.4256, corresponding to the mean of the annotators correlation among each other, as we did not have ``gold labels'' for the queries --- query--passage evaluations by the actual query creators \cite{bailey2008relevance}, or provided by trained evaluators with access to detailed query descriptions, as done by~\cite{thomas2023large}. According to Damessie et al.~\cite{damessie2017gauging}, crowd workers have a Cohen's Kappa correlation from 0.24 to 0.52; although above the average, our human annotators correlation falls within that interval, despite our expectation of a higher correlation.

\begin{figure}[ht]
\centering
\includegraphics[width=.9\textwidth]{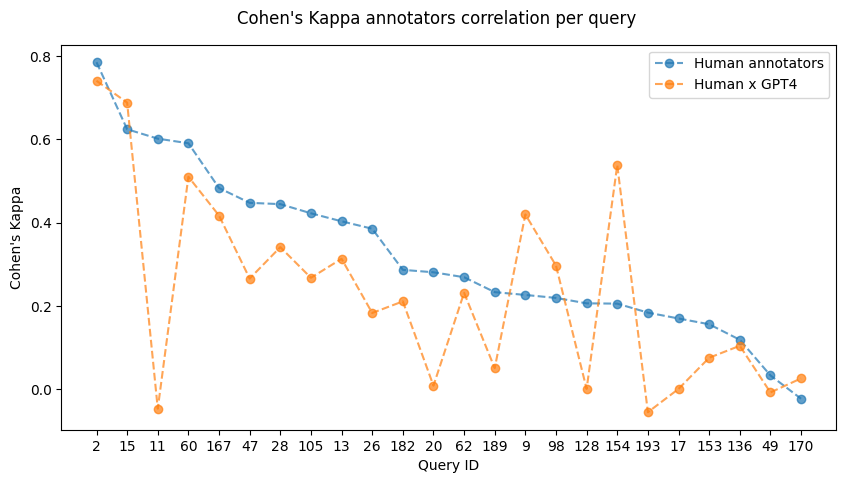}
\caption{\label{figure:annotators_correlation}When analyzed per query, most of the time humans and GPT-4 find the same questions more confusing to annotate for passage relevance, as the Cohen's Kappa correlation indicates. For 3 questions --- IDs 9, 98 and 154, GPT-4 got very correlated to Human Annotator 2, which explains the higher metrics for those questions.}
\end{figure}

Regardless of whether human or LLM, the annotators correlation varies per question --- inter-human annotation Cohen's Kappa varies from -0.0236 to 0.7857, while GPT-4 versus human annotators range from -0.0561 to 0.7411 --- and as shown in Figure \ref{figure:annotators_correlation} both humans and LLM find most of the time the same questions harder to evaluate for passage relevance --- harder meaning more confusing, hence reducing scores correlation. See Tables \ref{tab:cohen_kappa_per_question_part_01} and \ref{tab:cohen_kappa_per_question_part_02} in the Appendix for human versus human and human versus LLM correlations for each query.

Considering query--passage relevance annotation is a subjective task, we argue that a non-categorical metric would be more appropriate to measure the annotators' correlation, as errors by a single score level --- which indeed are very frequent for both humans and LLM annotators, as shown in the Appendix \ref{confusion_matrices_discussion} --- should be considered ``less critical'', or within the subjectivity interval which ends up being intrinsic for the task. Therefore, for completeness, we present the Spearman and Pearson correlations between humans and GPT-4 annotators in Tables \ref{tab:spearman_4_score} and \ref{tab:person_4_score}, respectively. Even though on both metrics human annotators' correlation is still well above their individual correlations with the GPT-4 scores, all the metrics are within a higher value, which we argue better captures the current LLM effectiveness on the query--passage relevance evaluation task, which is comparable to human crowd workers.

\begin{table*}
    \centering
    \renewcommand{\arraystretch}{1.2}
    \begin{tabular}{>{\raggedright}p{4.5cm}>{\centering}p{1.5cm}>{\centering}p{1.5cm}>{\centering}p{1.5cm}>{\centering\arraybackslash}p{1.5cm}}
        & \textbf{HA\textsubscript{1}} & \textbf{HA\textsubscript{2}} & \textbf{HA\textsubscript{3}} & \textbf{GPT-4}\\
        \midrule
        Human Annotator 1 (HA\textsubscript{1}) & --- & 0.4369 & 0.4294 & 0.3234\\
        Human Annotator 2 (HA\textsubscript{2}) & 0.4369 & --- & 0.4105 & 0.2593\\
        Human Annotator 3 (HA\textsubscript{3}) & 0.4294 & 0.4105 & --- & 0.3498\\
        Mean & \textbf{0.4331} & \textbf{0.4237} & \textbf{0.4199} & \textbf{0.3108}\\
        Standard Deviation & \textbf{0.0037} & \textbf{0.0132} & \textbf{0.0095} & \textbf{0.0380}\\
        Mean human annotators & \multicolumn{3}{c}{\textbf{0.4256\textpm0.0055}} & ---\\
        Diff. human annotators Mean & 0.0076 & -0.0019 & -0.0057 & -0.1096\\
        \bottomrule
    \end{tabular}
    \caption{Cohen's Kappa correlations for the 4-score evaluations.}
    \label{tab:cohen_4_score}
\end{table*}

\begin{table*}
    \centering
    \renewcommand{\arraystretch}{1.2}
    \begin{tabular}{>{\raggedright}p{4.5cm}>{\centering}p{1.5cm}>{\centering}p{1.5cm}>{\centering}p{1.5cm}>{\centering\arraybackslash}p{1.5cm}}
        & \textbf{HA\textsubscript{1}} & \textbf{HA\textsubscript{2}} & \textbf{HA\textsubscript{3}} & \textbf{GPT-4}\\
        \midrule
        Human Annotator 1 (HA\textsubscript{1}) & --- & 0.6931 & 0.6924 & 0.6073\\
        Human Annotator 2 (HA\textsubscript{2}) & 0.6931 & --- & 0.6985 & 0.6174\\
        Human Annotator 3 (HA\textsubscript{3}) & 0.6924 & 0.6985 & --- & 0.6296\\
        Mean & \textbf{0.6927} & \textbf{0.6958} & \textbf{0.6954} & \textbf{0.6181}\\
        Standard Deviation & \textbf{0.0004} & \textbf{0.0027} & \textbf{0.0031} & \textbf{0.0091}\\
        Mean human annotators & \multicolumn{3}{c}{\textbf{0.6946\textpm0.0014}} & ---\\
        Diff. human annotators Mean & -0.0019 & 0.0011 & 0.0008 & -0.0596\\
        \bottomrule
    \end{tabular}
    \caption{Spearman correlations for the 4-score evaluations.}
    \label{tab:spearman_4_score}
\end{table*}

\begin{table*}
    \centering
    \renewcommand{\arraystretch}{1.2}
    \begin{tabular}{>{\raggedright}p{4.5cm}>{\centering}p{1.5cm}>{\centering}p{1.5cm}>{\centering}p{1.5cm}>{\centering\arraybackslash}p{1.5cm}}
        & \textbf{HA\textsubscript{1}} & \textbf{HA\textsubscript{2}} & \textbf{HA\textsubscript{3}} & \textbf{GPT-4}\\
        \midrule
        Human Annotator 1 (HA\textsubscript{1}) & --- & 0.6982 & 0.6973 & 0.5982\\
        Human Annotator 2 (HA\textsubscript{2}) & 0.6982 & --- & 0.7132 & 0.6146\\
        Human Annotator 3 (HA\textsubscript{3}) & 0.6973 & 0.7132 & --- & 0.6326\\
        Mean & \textbf{0.6977} & \textbf{0.7057} & \textbf{0.7052} & \textbf{0.6151}\\
        Standard Deviation & \textbf{0.0004} & \textbf{0.0075} & \textbf{0.0079} & \textbf{0.0140}\\
        Mean human annotators & \multicolumn{3}{c}{\textbf{0.7029\textpm0.0037}} & ---\\
        Diff. human annotators Mean & -0.0051 & 0.0028 & 0.0024 & -0.0652\\
        \bottomrule
    \end{tabular}
    \caption{Pearson correlations for the 4-score evaluations.}
    \label{tab:person_4_score}
\end{table*}

The correlation between the relevance scores assigned by human annotators is higher than the correlation between humans and LLM. However, both have high variation depending on the question, and the overall correlation achieved by the LLM -- within the crowd workers correlation interval -- confirms the possibility of creating an evaluation dataset with a high quality/cost relation.

At the current development stage, LLM evaluation effectiveness is below human when distinguishing between closer categories ---``not relevant'' or ``on topic''; ``highly relevant'' or ``perfectly relevant'' --- but it is reasonable to expect that will improve, as LLMs improve their ability to understand text nuances. There is also room to enhance LLM passage relevance evaluation through additional prompt engineering, given LLM's sensitivity to the provided input \cite{thomas2023large}.

\subsection{Retrieval systems evaluation results}

\begin{table*}
    \centering
    \renewcommand{\arraystretch}{1.2}
    \begin{tabular}{>{\raggedright}p{7.0cm}>{\raggedleft\arraybackslash}p{1.7cm}}
        \textbf{Retrieval system} & \textbf{nDCG@10}\\
        \midrule
        \textbf{1M Passages} & \\
        BM25 & 0.3156\\
        E5-base & 0.3955\\
        ColBERT-X mMARCO pt-BR & 0.4039\\
        BM25 + E5-large & 0.4418\\
        text-embedding-3-small half & 0.4463 \\
        text-embedding-ada-002 & 0.4599\\
        text-embedding-3-small & 0.4635\\
        text-embedding-3-large & 0.5179\\
        BM25 + mT5-XL & 0.5310\\
        \midrule
        \textbf{10M Passages} & \\
        BM25 & 0.4467\\
        E5-large & 0.5563\\
        SPLADE v2 pt-BR & 0.5806\\
        E5-large + SPLADE v2 pt-BR RRF & 0.6272\\
        ColBERT-X mMARCO pt-BR & 0.6279\\
        BM25 + E5-large & 0.6364\\
        E5-large + ColBERT-X mMARCO pt-BR RRF & 0.6377\\
        SPLADE v2 pt-BR + mT5-XL & 0.6966\\
        BM25 + mT5-XL & 0.7109\\
        \bottomrule
    \end{tabular}
    \caption{The nDCG@10 effectiveness on the 50 test queries. The relevance scores are from the LLM for the 10M dataset.}
    \label{tab:retrievers_10M_effectiveness}
\end{table*}

\begin{table*}
    \centering
    \renewcommand{\arraystretch}{1.2}
    \begin{tabular}{>{\raggedright}p{7.0cm}>{\raggedleft\arraybackslash}p{1.7cm}}
        \textbf{Retrieval system} & \textbf{nDCG@10}\\
        \midrule
        BM25 & 0.3991\\
        ColBERT-X mMARCO pt-BR & 0.4927\\
        BM25 + E5-large & 0.5423\\
        text-embedding-ada-002 & 0.5630\\
        text-embedding-3-small & 0.5688\\
        text-embedding-3-large & 0.6319\\
        BM25 + mT5-XL & 0.6593\\
        \bottomrule
    \end{tabular}
    \caption{Retrieval systems nDCG@10 effectiveness on the 50 test queries, 1M dataset.}
    \label{tab:retrievers_1M_effectiveness}
\end{table*}

We evaluated the retrievers effectiveness using the LLM annotated query--passages (qrels): Tables \ref{tab:retrievers_10M_effectiveness} and \ref{tab:retrievers_1M_effectiveness} present the nDCG@10 effectiveness for the 10M and 1M datasets respectively. We included the evaluation of the 1M retrievers runs using the 10M qrels, as can be seen in Table \ref{tab:retrievers_10M_effectiveness}, to emphasize the stronger retrievers effectiveness, which do a better job finding the remaining relevant passages in the 1M dataset version, than weaker ones running over the entire 10M dataset. BM25 + mT5 XL retriever illustrates that: it held the best effectiveness not only over both datasets, but also when being evaluated against the 10M qrels over the results of a 1M run. Due to the corresponding costs, we ran the commercial retrievers only against the 1M dataset.

The ranking of retrievers with respect to effectiveness matched our expectations, following the literature. We consider that an additional indication of the overall datasets quality as, despite being created in a semi-automated cost-effective way, they are able to evaluate a diversity of retrievers.

\section{Conclusion}

We introduced the Quati dataset to support the development of IR systems for Brazilian Portuguese retrieval tasks. Quati is publicly available in two sizes, 10M a 1M passages, with 50-query qrels with respectively an average of 97.78 and 38.66 annotated passages per query.
Through comparisons with human annotators, we show that state-of-the-art LLM can be used in a semi-automated and cost-effective way to create IR datasets for a specific target language, in the query--passage annotation role, with equivalent performance of humans: LLM annotations correlate with humans' in similar way human crowd workers annotations do. We showed the correlation strongly varies per question for both human--human or LLM--human comparison, and the overall annotations are comparable to the result of human crowd workers, largely adopted for dataset creation, for a fraction of the cost. 

Although we aimed at creating high-quality evaluation datasets, we understand that is only part of supporting the development of IR systems. We intent to expand this work creating a training dataset, and assess the impact on effectiveness of training different IR systems with datasets specifically created for the target language. We also want to further explore the LLM prompt development and quantify its impact over the LLM--human annotations correlation.

Finally, we open-sourced the generation scripts encouraging the dataset expansion with further annotations, or the generation of information retrieval datasets targeting other languages.

\section*{Acknowledgements} We thank Leodécio Braz da Silva Segundo for the valuable support during the human annotation task. We also thank Leonardo Benardi de Avila and Monique Monteiro for the SPLADE v2 retrievals, using the model they trained for Brazilian Portuguese. This research was partially funded by grant 2022/01640-2 from Fundação de Amparo à Pesquisa do Estado de São Paulo (FAPESP).

%
%

\bibliographystyle{plain}
\bibliography{main.bib}

\clearpage
\newpage

\appendix

\section{Appendix}

\subsection{Human Annotators confusion matrices}\label{confusion_matrices_discussion}

Tables \ref{tab:annotator01_LLM_confusion_matrix}, \ref{tab:annotator02_LLM_confusion_matrix}, and \ref{tab:annotator03_LLM_confusion_matrix} present the confusion matrix between each human annotator and GPT-4 illustrating the LLM behavior. Similar to the findings of Faggioli et al.~\cite{faggioli2023perspectives}, we verified GPT-4 tends to evaluate the passages with higher scores when compared to humans. For instance, when compared to Human Annotator 1, GPT-4 classified as ``Relevant (1)'' 13 of 52 (25\%) query--passages annotated as ``Irrelevant (0)'' by the human; that was the higher misclassification on that score. A similar behavior is verified for the other scores, with GPT-4 indicating as ``Highly relevant (2)'' 18 of 68 (26.47\%) ``Relevant (1)'', or misclassifying as ``Perfectly relevant (3)'' 27 of 65 (41,54\%) ``Highly relevant (2)'' query--passages. In fewer cases, GPT-4 classifies as ``Irrelevant (0)'' something humans have annotated as either ``Highly relevant (2)'' or ``Perfectly relevant (3)'' or vice-versa.

The human annotators presented similar behavior (Tables \ref{tab:annotator01_annotator02_confusion_matrix}, \ref{tab:annotator01_annotator03_confusion_matrix}, and \ref{tab:annotator02_annotator03_confusion_matrix}), but disagreeing more on the intermediate scores --- ``Relevant (1)'' and ``Highly relevant (2)'' --- indicating they also suffer from the boundary uncertainty when determining the query--passage relevance score.

\begin{table*}
    \centering
    \renewcommand{\arraystretch}{1.4}
    \begin{tabular}{>{\raggedleft}p{0.8cm}>{\raggedleft}p{3.0cm}>{\raggedleft}p{1.cm}>{\raggedleft}p{1.cm}>{\raggedleft}p{1.cm}>{\raggedleft}p{1.cm}>{\raggedleft\arraybackslash}p{1.5cm}}
        & & \multicolumn{4}{c}{\textbf{GPT-4}} & \multirow{2}{1.5cm}{\raggedleft{}\textbf{Human totals}}\\
        & \textbf{Score} & \textbf{0} & \textbf{1} & \textbf{2} & \textbf{3} & \\
        \midrule
        \multirow{4}{*}{\rotatebox[origin=c]{90}{\parbox{2.1cm}{
            \centering 
            \textbf{Human Annotator 1}
            \vspace{5pt}}}
        }
        & Irrelevant (0)         & 25 & 13 & 12 &  2 & 52\\
        & Relevant (1)           & 12 & 24 & 18 & 14 & 68\\
        & Highly relevant (2)    &  4 & 11 & 23 & 27 & 65\\
        & Perfectly relevant (3) &  1 &  5 &  3 & 46 & 55\\
        \midrule
        & \textbf{GPT-4 totals}  & 42 & 53 & 56 & 89 & \\
        \bottomrule
    \end{tabular}
    \caption{Human Annotator 1 and GPT-4 scores confusion matrix.}
    \label{tab:annotator01_LLM_confusion_matrix}
\end{table*}

\begin{table}[H]
    \centering
    \renewcommand{\arraystretch}{1.4}
    \begin{tabular}{>{\raggedleft}p{0.8cm}>{\raggedleft}p{3.0cm}>{\raggedleft}p{1.cm}>{\raggedleft}p{1.cm}>{\raggedleft}p{1.cm}>{\raggedleft}p{1.cm}>{\raggedleft\arraybackslash}p{1.5cm}}
        & & \multicolumn{4}{c}{\textbf{GPT-4}} & \multirow{2}{1.5cm}{\raggedleft{}\textbf{Human totals}}\\
        & \textbf{Score} & \textbf{0} & \textbf{1} & \textbf{2} & \textbf{3} & \\
        \midrule
        \multirow{4}{*}{\rotatebox[origin=c]{90}{\parbox{2.1cm}{
            \centering 
            \textbf{Human Annotator 2}
            \vspace{5pt}}}
        }
        & Irrelevant (0)         & 29 & 17 & 11 &  2 & 59\\
        & Relevant (1)           &  5 & 15 & 17 & 10 & 47\\
        & Highly relevant (2)    &  8 & 20 & 24 & 40 & 92\\
        & Perfectly relevant (3) &  0 &  1 &  4 & 37 & 42\\
        \midrule
        & \textbf{GPT-4 totals}  & 42 & 53 & 56 & 89 & \\
        \bottomrule
    \end{tabular}
    \caption{Human Annotator 2 and GPT-4 scores confusion matrix.}
    \label{tab:annotator02_LLM_confusion_matrix}
\end{table}

\begin{table}[H]
    \centering
    \renewcommand{\arraystretch}{1.4}
    \begin{tabular}{>{\raggedleft}p{0.8cm}>{\raggedleft}p{3.0cm}>{\raggedleft}p{1.cm}>{\raggedleft}p{1.cm}>{\raggedleft}p{1.cm}>{\raggedleft}p{1.cm}>{\raggedleft\arraybackslash}p{1.5cm}}
        & & \multicolumn{4}{c}{\textbf{GPT-4}} & \multirow{2}{1.5cm}{\raggedleft{}\textbf{Human totals}}\\
        & \textbf{Score} & \textbf{0} & \textbf{1} & \textbf{2} & \textbf{3} & \\
        \midrule
        \multirow{4}{*}{\rotatebox[origin=c]{90}{\parbox{2.1cm}{
            \centering 
            \textbf{Human Annotator 3}
            \vspace{5pt}}}
        }
        & Irrelevant (0)         & 29 & 13 & 10 &  1 & 53\\
        & Relevant (1)           &  7 & 21 & 11 & 13 & 52\\
        & Highly relevant (2)    &  6 & 12 & 19 & 19 & 56\\
        & Perfectly relevant (3) &  0 &  7 & 16 & 56 & 79\\
        \midrule
        & \textbf{GPT-4 totals}  & 42 & 53 & 56 & 89 & \\
        \bottomrule
    \end{tabular}
    \caption{Human Annotator 3 and GPT-4 scores confusion matrix. Best agreement on the highest score.}
    \label{tab:annotator03_LLM_confusion_matrix}
\end{table}

\begin{table}[H]
    \centering
    \renewcommand{\arraystretch}{1.4}
    \begin{tabular}{>{\raggedleft}p{1.2cm}>{\raggedleft}p{3.0cm}>{\raggedleft}p{1.cm}>{\raggedleft}p{1.cm}>{\raggedleft}p{1.cm}>{\raggedleft}p{1.cm}>{\raggedleft\arraybackslash}p{1.5cm}}
        & & \multicolumn{4}{c}{\textbf{Human Annotator 2 (HA\textsubscript{2})}} & \multirow{2}{1.5cm}{\raggedleft{}\textbf{HA\textsubscript{1} totals}}\\
        & \textbf{Score} & \textbf{0} & \textbf{1} & \textbf{2} & \textbf{3} & \\
        \midrule
        \multirow{4}{*}{\rotatebox[origin=c]{90}{\parbox{2.2cm}{
            \centering 
            \textbf{Human Annotator 1 (HA\textsubscript{1})}
            \vspace{5pt}}}
        }
        & Irrelevant (0)             & 41 &  6 &  4 &  1 & 52\\
        & Relevant (1)               & 13 & 25 & 28 &  2 & 68\\
        & Highly relevant (2)        &  4 & 11 & 42 &  8 & 65\\
        & Perfectly relevant (3)     &  1 &  5 & 18 & 31 & 55\\
        \midrule
        & \textbf{HA\textsubscript{2} totals} & 59 & 47 & 92 & 42 & \\
        \bottomrule
    \end{tabular}
    \caption{Human Annotator 1 and Human Annotator 2 scores confusion matrix. There is no clear disagreement tendency (lower or higher score).}
    \label{tab:annotator01_annotator02_confusion_matrix}
\end{table}

\begin{table}[H]
    \centering
    \renewcommand{\arraystretch}{1.4}
    \begin{tabular}{>{\raggedleft}p{1.2cm}>{\raggedleft}p{3.0cm}>{\raggedleft}p{1.cm}>{\raggedleft}p{1.cm}>{\raggedleft}p{1.cm}>{\raggedleft}p{1.cm}>{\raggedleft\arraybackslash}p{1.5cm}}
        & & \multicolumn{4}{c}{\textbf{Human Annotator 3 (HA\textsubscript{3})}} & \multirow{2}{1.5cm}{\raggedleft{}\textbf{HA\textsubscript{1} totals}}\\
        & \textbf{Score} & \textbf{0} & \textbf{1} & \textbf{2} & \textbf{3} & \\
        \midrule
        \multirow{4}{*}{\rotatebox[origin=c]{90}{\parbox{2.2cm}{
            \centering 
            \textbf{Human Annotator 1 (HA\textsubscript{1})}
            \vspace{5pt}}}
        }
        & Irrelevant (0)             & 41 &  7 &  1 &  3 & 52\\
        & Relevant (1)               &  9 & 29 & 21 &  9 & 68\\
        & Highly relevant (2)        &  2 & 11 & 26 & 26 & 65\\
        & Perfectly relevant (3)     &  1 &  5 &  8 & 41 & 55\\
        \midrule
        & \textbf{HA\textsubscript{3} totals} & 53 & 52 & 56 & 79 & \\
        \bottomrule
    \end{tabular}
    \caption{Human Annotator 1 and Human Annotator 3 scores confusion matrix.}
    \label{tab:annotator01_annotator03_confusion_matrix}
\end{table}

\begin{table}[H]
    \centering
    \renewcommand{\arraystretch}{1.4}
    \begin{tabular}{>{\raggedleft}p{1.2cm}>{\raggedleft}p{3.0cm}>{\raggedleft}p{1.cm}>{\raggedleft}p{1.cm}>{\raggedleft}p{1.cm}>{\raggedleft}p{1.cm}>{\raggedleft\arraybackslash}p{1.5cm}}
        & & \multicolumn{4}{c}{\textbf{Human Annotator 3 (HA\textsubscript{3})}} & \multirow{2}{1.5cm}{\raggedleft{}\textbf{HA\textsubscript{2} totals}}\\
        & \textbf{Score} & \textbf{0} & \textbf{1} & \textbf{2} & \textbf{3} & \\
        \midrule
        \multirow{4}{*}{\rotatebox[origin=c]{90}{\parbox{2.2cm}{
            \centering 
            \textbf{Human Annotator 2 (HA\textsubscript{2})}
            \vspace{5pt}}}
        }
        & Irrelevant (0)             & 44 & 11 &  1 &  3 & 59\\
        & Relevant (1)               &  5 & 19 & 14 &  9 & 47\\
        & Highly relevant (2)        &  4 & 21 & 35 & 32 & 92\\
        & Perfectly relevant (3)     &  0 &  1 &  6 & 35 & 42\\
        \midrule
        & \textbf{HA\textsubscript{3} totals} & 53 & 52 & 56 & 79 & \\
        \bottomrule
    \end{tabular}
    \caption{Human Annotator 2 and Human Annotator 3 scores confusion matrix.}
    \label{tab:annotator02_annotator03_confusion_matrix}
\end{table}

\subsection{Annotators correlation per question}

\begin{table}[H]
    \centering
    \renewcommand{\arraystretch}{1.1}
    \begin{tabular}{>{\raggedright}p{4.5cm}>{\raggedleft}p{1.6cm}>{\raggedleft}p{1.6cm}>{\raggedleft}p{1.6cm}>{\raggedleft\arraybackslash}p{2.5cm}}
        \multirow{2}{*}{\textbf{Query}} & \textbf{a1\texttimes{}a2} & \textbf{a2\texttimes{}a3} & \textbf{a3\texttimes{}a1} & \multirow{2}{*}{\textbf{Mean\textpm{}Std}} \\
        & \textbf{a1\texttimes{}LLM} & \textbf{a2\texttimes{}LLM} & \textbf{a3\texttimes{}LLM} & \\
        \midrule

        \multirow{2}{5.0cm}{Onde está localizada a Praça XV de Novembro?}
        & 0.2647 & 0.6970 & 0.3056 & 0.4224\textpm0.1949\\
        & 0.2857 & 0.2105 & 0.3056 & 0.2673\textpm0.0409\\
        
        \midrule
        \multirow{2}{5.0cm}{Qual foi a importância da usina de Volta Redonda RJ para a industrialização brasileira?} 
        & -0.0127 & 0.1026 & 0.2647 & 0.1182\textpm0.1138\\
        &  0.0909 & 0.1176 & 0.1026 & 0.1037\textpm0.0109\\[0.3cm]

        \midrule
        \multirow{2}{5.0cm}{Qual o uso dos códigos SWIFT?} 
        & 0.6154 & 0.0000 & 0.0000 & 0.2051\textpm0.2901\\
        & 1.0000 & 0.6154 & 0.0000 & 0.5385\textpm0.4119\\

        \midrule
        \multirow{2}{5.0cm}{O que são os celulares ``mid-range''?} 
        & 0.5082 & 0.2537 & 0.6875 & 0.4831\textpm0.1780\\
        & 0.5238 & 0.2188 & 0.5082 & 0.4169\textpm0.1403\\

        \midrule
        \multirow{2}{5.0cm}{Por que os países Guiana e Suriname não são filiados a Conmebol?} 
        & 0.8361 & 0.8438 & 0.6774 & 0.7857\textpm0.0767\\
        & 0.6825 & 0.6970 & 0.8438 & 0.7411\textpm0.0728\\[0.3cm]

        \midrule
        \multirow{2}{5.0cm}{quais os critérios de definição dos monumentos intitulados maravilhas do mundo moderno?} 
        & 0.7015 & 0.5455 &  0.5588 &  0.6019\textpm0.0706\\
        & 0.0278 & 0.0141 & -0.1842 & -0.0474\textpm0.0969\\[0.3cm]

        \midrule
        \multirow{2}{5.0cm}{Qual a maior torcida de futebol do Brasil?} 
        & 0.8077 & 0.6429 & 0.4231 & 0.6245\textpm0.1576\\
        & 0.4231 & 0.6429 & 1.0000 & 0.6886\textpm0.2377\\

        \midrule
        \multirow{2}{5.0cm}{Quando se realizou o plebiscito popular para definir o sistema político do Brasil?} 
        & 0.0000 & 0.0000 & 0.5082 & 0.1694\textpm0.2396\\
        & 0.0000 & 0.0000 & 0.0000 & 0.0000\textpm0.0000\\[0.3cm]

        \midrule
        \multirow{2}{5.0cm}{Como transformar uma cidade pacata em um polo turístico?} 
        & 0.2857 & 0.5833 & 0.4737 & 0.4476\textpm0.1229\\
        & 0.4030 & 0.1250 & 0.2683 & 0.2654\textpm0.1135\\

        \midrule
        \multirow{2}{5.0cm}{Quais são os melhores parques nacionais de Portugal?} 
        & -0.0811 & 0.4286 & -0.2500 &  0.0325\textpm0.2884\\
        & -0.4286 & 0.1304 &  0.2727 & -0.0085\textpm0.3027\\

        \midrule
        \multirow{2}{5.0cm}{Quando foi criada a consolidação das leis trabalhistas no brasil?} 
        & 0.5833 & 0.5946 & 0.5946 & 0.5908\textpm0.0053\\
        & 0.4595 & 0.5714 & 0.5000 & 0.5103\textpm0.0463\\

        \midrule
        \multirow{2}{5.0cm}{Quais partidos já ocuparam o cargo da presidência do Brasil?} 
        & -0.0448 & 0.0000 & 0.8507 & 0.2687\textpm0.4120\\
        &  0.2424 & 0.3333 & 0.1176 & 0.2311\textpm0.0884\\

        \bottomrule
           
    \end{tabular}
    \caption{Annotators Cohen's Kappa correlation per question, for the initial 12 questions. For each question, first line contains the Human Annotators correlation, second line the Human x GPT-4 correlation. Zeroed correlations for two or more different annotators for a same question is due to one of the annotators have applied the same score for all the 10 passages of that given query.}
    \label{tab:cohen_kappa_per_question_part_01}
\end{table}

\begin{table}[H]
    \centering
    \renewcommand{\arraystretch}{1.1}
    \begin{tabular}{>{\raggedright}p{4.5cm}>{\raggedleft}p{1.6cm}>{\raggedleft}p{1.6cm}>{\raggedleft}p{1.6cm}>{\raggedleft\arraybackslash}p{2.5cm}}
        \multirow{2}{*}{\textbf{Query}} & \textbf{a1\texttimes{}a2} & \textbf{a2\texttimes{}a3} & \textbf{a3\texttimes{}a1} & \multirow{2}{*}{\textbf{Mean\textpm{}Std}} \\
        & \textbf{a1\texttimes{}LLM} & \textbf{a2\texttimes{}LLM} & \textbf{a3\texttimes{}LLM} & \\
        \midrule

        \multirow{2}{5.0cm}{Quando é o dia Mundial da Alimentação?} 
        & 0.4737 & 0.2308 & -0.0870 & 0.2058\textpm0.2296\\
        & 0.0000 & 0.0000 &  0.0000 & 0.0000\textpm0.0000\\
        \midrule

        \multirow{2}{5.0cm}{Quais os tipos de denominação (DO) que os vinhos podem receber?}
        & 0.2857 &  0.1111 & 0.0698 & 0.1555\textpm0.0936\\
        & 0.2593 & -0.0345 & 0.0000 & 0.0749\textpm0.1311\\[0.3cm]
        
        \midrule
        \multirow{2}{5.0cm}{Quais as causas para lábios inflamados em crianças?} 
        & 0.1803 & -0.2121 & -0.0390 & -00236\textpm0.1606\\
        & 0.0000 & -0.1667 &  0.2424 & 0.0253\textpm0.1680\\

        \midrule
        \multirow{2}{5.0cm}{Quais as origens de pessoas com olhos verdes?} 
        & 0.4595 & 0.2105 & 0.1892 & 0.2864\textpm0.1227\\
        & 0.3421 & 0.2105 & 0.0789 & 0.2105\textpm0.1074\\

        \midrule
        \multirow{2}{5.0cm}{No que se difere o civismo da cidadania?} 
        & 0.1566 & 0.3750 & 0.1667 & 0.2328\textpm0.1007\\
        & 0.0909 & 0.0588 & 0.0000 & 0.0499\textpm0.0376\\

        \midrule
        \multirow{2}{5.0cm}{Quais atitudes podem prejudicar a saúde mental?} 
        & -0.0606 &  0.0278 & 0.5833 &  0.1835\textpm0.2850\\
        & -0.0465 & -0.1494 & 0.0278 & -0.0561\textpm0.0727\\

        \midrule
        \multirow{2}{5.0cm}{quais países europeus seguem o regime monarquista?} 
        & 0.0909 & 0.4643 & 0.1228 & 0.2260\textpm0.1690\\
        & 0.2424 & 0.7059 & 0.3103 & 0.4196\textpm0.2044\\

        \midrule
        \multirow{2}{5.0cm}{Como o Brasil reagiu a epidemia de AIDS no fim do século XX?} 
        & 0.5161 & 0.4286 & 0.2647 & 0.4031\textpm0.1042\\
        & 0.4366 & 0.1781 & 0.3243 & 0.3130\textpm0.1059\\

        \midrule
        \multirow{2}{5.0cm}{Por que a legislação de um país é tão importante?} 
        & 0.1667 &  0.1667 &  0.5082 & 0.2805\textpm0.1610\\
        & 0.1176 & -0.0417 & -0.0526 & 0.0078\textpm0.0778\\

        \midrule
        \multirow{2}{5.0cm}{Como podemos classificar o relevo brasileiro?} 
        & 0.3750 & 0.2647 & 0.5161 & 0.3853\textpm0.1029\\
        & 0.1667 & 0.0789 & 0.3023 & 0.1826\textpm0.0919\\

        \midrule
        \multirow{2}{5.0cm}{Existem vantagens ao definir uma moeda única?} 
        & 0.7222 & 0.3056 & 0.3056 & 0.4444\textpm0.1964\\
        & 0.3056 & 0.3056 & 0.4118 & 0.3410\textpm0.0501\\

        \midrule
        \multirow{2}{5.0cm}{Qual o critério para classificação para a Copa do Brasil?} 
        & 0.2405 & 0.3750 & 0.0411 & 0.2189\textpm0.1372\\
        & 0.1250 & 0.5000 & 0.2647 & 0.2966\textpm0.1547\\

        \bottomrule
           
    \end{tabular}
    \caption{Annotators Cohen's Kappa correlation per question, for the last 12 questions. As in Table \ref{tab:cohen_kappa_per_question_part_01}, for each listed question, the first line contains the Human Annotators correlation, second line the Human x GPT-4 correlation. Also, Zeroed correlations for two or more different annotators for the same question are due to one of the annotators having applied the same score for all the 10 passages of that given query.}
    \label{tab:cohen_kappa_per_question_part_02}
\end{table}

\end{document}